%% file: Labiano.tex
\documentclass{elsart3}

\usepackage{natbib}

\usepackage{graphics}
 \usepackage{graphicx}
 \usepackage{epsfig}

\usepackage{amssymb}

\begin{document}

\begin{frontmatter}



\title{Star formation in hosts of young radio galaxies}


\author{A. Labiano$^a$},
\author{C. P. O'Dea$^b$},
\author{P. D. Barthel$^a$},
\author{W. H. de Vries$^c$},
\author{S. A. Baum$^d$}

\address[]{Kapteyn Astronomical Institute, Groningen, 9700 AV, The Netherlands.}
\address[]{Department of Physics, Rochester Institute of Technology, Rochester, NY, 14623, USA.}
\address[]{Lawrence Livermore National Lab., Livermore CA, 94550, USA.}
\address[]{Center for Imaging Science, Rochester Institute of Technology,  Rochester, NY 14623. USA}

\begin{abstract}
% Text of abstract
We present near ultraviolet imaging with the Hubble Space Telescope Advanced Camera for 
Surveys, targeting young radio galaxies (Gigahertz Peaked Spectrum and Compact Steep 
Spectrum sources), in search of star formation regions in their hosts. 
We find near UV light which could be the product of recent star formation in eight of 
the nine observed sources. However, observations at other wavelengths and colors 
are needed to definitively establish the nature of the observed UV light. 
In the CSS sources 1443+77 and 1814--637 the near UV light is aligned with and is co-spatial
with the radio source, and we suggest that in these sources the UV light is
produced by star formation triggered and/or enhanced by the radio source.  
%We present preliminary results on the UV properties of the sources and look for star formation regions in the hosts. We find possible traces of recent star formation in eight of the nine observed sources. However, observations at other wavelengths and colors are needed to further asses the nature of the observed UV properties. CSS sources 1443+77 and 1814--637 show star formation enhanced by the expansion of the radio lobes through the host.
%The near UV emission is very sensitive to the presence of hot young stars and therefore recent star formation events. 
%We compare the UV properties of GPS, CSS, and large scale radio sources and study star formation history as a function of the relative age of the radio source, and its relation with the expanding radio source. Stellar synthesis models suggest that the host has undergone an burst of star formation not much sooner than the formation of the radio source. 

\end{abstract}

\begin{keyword}
%Galaxies: elliptical
Galaxies: stellar content
\sep Galaxies: photometry
\sep Ultraviolet: galaxies



\end{keyword}

\end{frontmatter}

% main text

\section{Introduction}

The relationship between black hole mass and galaxy mass implies that the growth and evolution 
of black holes (therefore AGN) and their host galaxies must somehow be related 
\citep[e.g.,][]{Gebhardt00}. Mergers and strong interactions can trigger AGN activity in 
a galaxy \citep[e.g.,][]{Heckman86, Baum92, Israel98}. These events can also produce 
instabilities in the ISM and trigger star formation (e.g., Ho 2005). Numerical simulations 
\citep[e.g.,][]{Mellema02,Rees89} suggest that the advancement of the jets through 
the host galaxy environment can also trigger star formation. UV studies of large 
3CR sources find traces of episodes of star formation around the time when the radio 
source was triggered \citep[i.e. $\lesssim 10^7 - 10^8$~yr,][]{Koekemoer99, Allen02, O'Dea01, O'Dea03, Martel02} suggesting a possible link between both. % Chiaberge02,

Gigahertz Peaked Spectrum (GPS) sources and Compact Steep Spectrum (CSS) are young, smaller \citep[GPS $\lesssim 1$ kpc, CSS $\lesssim$15 kpc, for a review see][]{O'Dea98} versions of the large powerful radio sources, so they are expected to exhibit signs of more recent star formation. Their size makes them excellent probes of the interactions between the expanding lobes and the host. They have not yet completely broken through the host ISM, so these interactions are expected to be even more important than in the larger sources.

The near UV observations are very sensitive to the presence of hot young stars and therefore will trace recent star formation events. We have obtained high resolution HST/ACS UV images of these young compact sources to study their morphology and the extent of recent star formation. %We also look for possible links between the AGN activity and the star formation history of the host.

Our sample is chosen to be representative of GPS and CSS sources with z$\lesssim$ 0.5, nearby enough to eliminate strong effects due to evolution with cosmic time. The objects are drawn primarily from the well-defined samples of \citet{Fanti90, Fanti01} and \citet{Stanghellini97}. \\  %The complete sample (25 sources) is being observed through a currently ongoing snapshot program. The sub-sample presented here consists of the nine sources observed at the time of writing. % The comparison sample of 3CR sources consists of FR 1 and FR 2 sources with redshifts less than 0.1 observed near-UV by \citet{Allen02}.

\section{Observations and data reduction}

We obtained high resolution near-UV snapshot images with the High Resolution Channel (HRC) of the Advanced Camera for surveys (ACS) on board the Hubble Space Telescope, using the F330W filter. The objects observed are GPS and CSS galaxies 1117+146, 1233+418, 1345+125, 1443+77, 1607+268, 1814--637, 1934--638, 1946+708, 2352+495. Here we present preliminary results on the UV properties of these sources.
%present the preliminary results on the most interesting sources.
%These observations are part of a bigger program of snapshot ACS observations of 25 GPS and CSS galaxies. 

The two 2-D fitting code GALFIT \citep{Peng02} was used to parameterize the UV emission. For each image we tested different combinations of point source and Sersic profiles, allowing the sky level, position and magnitudes of all components, as well the index and effective radii of the Sersic components, to vary. The final model was chosen according to the lowest $\chi^2$ and best residuals (with the lowest number of components). There are no good models of the Point Spread Function (PSF) for the ACS/HRC. To model the PSF we used the {\it calibration plan} observations of Cycles 12 and 13.

The observed objects are Narrow Line Radio Galaxies (NLRG) so we expect no contamination from the AGN nucleus. The main contribution from emission line gas to our observations would come from MgII. It is usually only found in the nuclear Broad Line Region (BLR) of AGN hosts so we do not expect contamination from emission line gas either.

\section{Results}

\input{TABLES/ACSPosi.tex}

The summary of the GALFIT models is shown in Table \ref{ACSPosi}. The UV images together with GALFIT models are shown in Figure \ref{galfigs}. 

All sources consist of at least two components except 1117+146 (a point source). The hosts of 1233+418, 1345+125, 1443+77, 1607+268, 1814--637 and 1934--638 show a combination of at least one Sersic component (with different indices) and one or several point sources. 1946+708 and 2352+495 show a combination of two and three point sources respectively\footnote{The Sersic profiles are used to parameterize the data, It does not necessarily imply that these UV components are galaxies.}.

%WE ARE USING THE SERSIC PROFILES TO PARAMETRIZE THE DATA, . WE ARE NOT NECESSARILY SUGGESTING TAT THESE UV COMPONENTS ARE GALAXIES.

The two brightest sources in the UV, 1443+77 and 1814--637, show alignment of the UV emission with the radio source \citep[see VLBI maps in][]{Tzioumis02, Sanghera95}. Both sources are CSS with radio sizes comparable to the separation between the UV components ($\sim$7 and $\sim0.5$ kpc respectively). The similarity in radio and UV sizes, alignment and higher UV emission suggest that the expansion of the radio source is enhancing star formation in the hosts of 1443+77 and 1814--637. \citet{Labiano05} found the presence of gas ionized by the shocks from the radio source in CSS sources. Furthermore, they find that 1443+77 shows the strongest contribution from shocks. We could expect these shocks to be affecting the star formation in the host too.

\input{FIGS/Galfig.tex}
\section{Summary}

We have obtained HST/ACS near-UV high resolution images of young radio sources: 
GPS and CSS galaxies. We detect near UV emission (point sources and/or  clumps) in
eight of the sources, consistent with the presence of recent star formation. 
In two CSS sources, 1443+77 and 1814--637 the near UV emission is aligned with and 
co-spatial with the the radio emission and we suggest that star formation has been
triggered/enhanced by  expansion of the radio source through the host. 
Observations at other wavelengths and measurement of the colors are needed to further 
asses the nature of the observed UV properties.

%The higher luminosity and alignment in 1443+77 and 1814--637 suggest that the host is increasing his star formation due to the expansion of the radio source. that we are seeing star formation and UV emission enhanced by the radio source expansion. 
%Stellar synthesis models suggest that the observed luminosities are due to a star formation burst burst of $\sim10^7$M$\odot$ about 1 or 10 Myr ago, implying that the hosts have undergone a burst of star formation not much sooner than the formation of the radio source.

%The connection between the AGN and the star formation is not yet clear but our observations suggest that some connection exists. The bursts of star formation could have triggered or at least favored the presence of the radio source, and in some sources, the expansion of the lobes is enhancing star formation in the host. 

%\section{}
%\label{}

% \appendix

% \section{}
% \label{}

% Bibliographic references with the natbib package:
% Parenthetical: \citep{Bai92} produces (Bailyn 1992).
% Textual: \citet{Bai95} produces Bailyn et al. (1995).
% An affix and part of a reference:
%   \citep[e.g.][Ch. 2]{Bar76}
%   produces (e.g. Barnes et al. 1976, Ch. 2).

\newcommand\aj{AJ}
\newcommand\apj{ApJ}
\newcommand\apjl{ApJ}
\newcommand\apjs{ApJS}
\newcommand\mnras{MNRAS}
\newcommand\aaps{A\&AS}
\newcommand\aap{A\&A}
\newcommand\nat{Nature}
\newcommand\aapr{A\&A~Rev.}
\newcommand\pasp{PASP}

\end{document}

%% file: TABLES/ACSPosi.tex
%%%%% TABLE 1 %%%%%

\begin{table}[t]
\caption{The magnitudes are observed, not corrected from galactic extinction. R$_e$ is the effective radius in miliarcsec except for 1443+77 (in arcsec).The 3$\sigma$ detection limit for a point source (FWHM $\sim$3 pixels) is 25.8. }
\label{ACSPosi}
\begin{minipage}{\columnwidth}
\centering
%\resizebox{\textwidth}{!}{
\begin{tabular}{ccccc}
\hline
\hline
%  &  \multicolumn{2}{c}{Radio (J2000)} & \multicolumn{2}{c}{Optical (J2000)} \\
%\cline{2-3}  
%\cline{4-5}
Source   &  Component &  STMAG & R$_e$ & Sersic index  \\
\hline 
1117+146 	& Point source   & 24.06$\pm$0.06 \\

1233+418 	& Point source   & 25.43$\pm$0.16 \\
		 	& Sersic profile & 24.21$\pm$0.36 & 100$\pm$800 & 0.04$\pm$0.6\\
			
1345+125 	& Point source   & 23.77$\pm$0.05 \\
		 	& Sersic profile & 21.14$\pm$0.02 & 35 $\pm$0.5 & 0.33$\pm$0.03\\
		 	& Sersic profile & 21.20$\pm$0.04  & 108 $\pm$7 & 1.62 $\pm$0.13\\
		 	& Sersic profile & 21.84$\pm$0.12  & 299 $\pm$58 & 2.29$\pm$0.38\\

1443+77   	& Point source   & 25.21 $\pm$0.10 \\
		 	& Point source   & 24.06 $\pm$0.06 \\
		 	& Sersic profile & 20.36 $\pm$0.16 & 1.1"$\pm$0.2" & 2.66$\pm$0.33\\ 

1607+268 	& Point source   & 24.64 $\pm$0.18\\
		 	& Sersic profile & 23.13 $\pm$0.14 &47$\pm$11 & 3.92$\pm$2.35 \\

1814--637	& Point source   & 16.95$\pm$0.10 \\
		 	& Sersic profile & 17.11 $\pm$0.12 & 25$\pm$4 & 2.08$\pm$0.31  \\

1934--638	& Point source   & 23.81$\pm$0.73 \\
		 	& Sersic profile & 21.75$\pm$0.12 & 39$\pm$5 &1.02$\pm$0.33\\
		 	& Sersic profile & 21.08$\pm$0.07 & 354$\pm$45&2.51$\pm$0.26\\

1946+708 	& Point source   & 25.81$\pm$0.25 \\
		 	& Point source   & 24.59$\pm$0.09 \\

2352+495 	& Point source   & 25.13$\pm$0.16 \\
		 	& Point source   & 25.53$\pm$0.19 \\
		 	& Point source   & 24.79$\pm$0.10 \\
\hline
\end{tabular}
%}
\end{minipage}
\end{table}
%%%%%%%%%%%%%%%

%% file: FIGS/Galfig.tex
\begin{figure*}[t]
\centering
\begin{center}
\includegraphics[width=0.45\columnwidth]{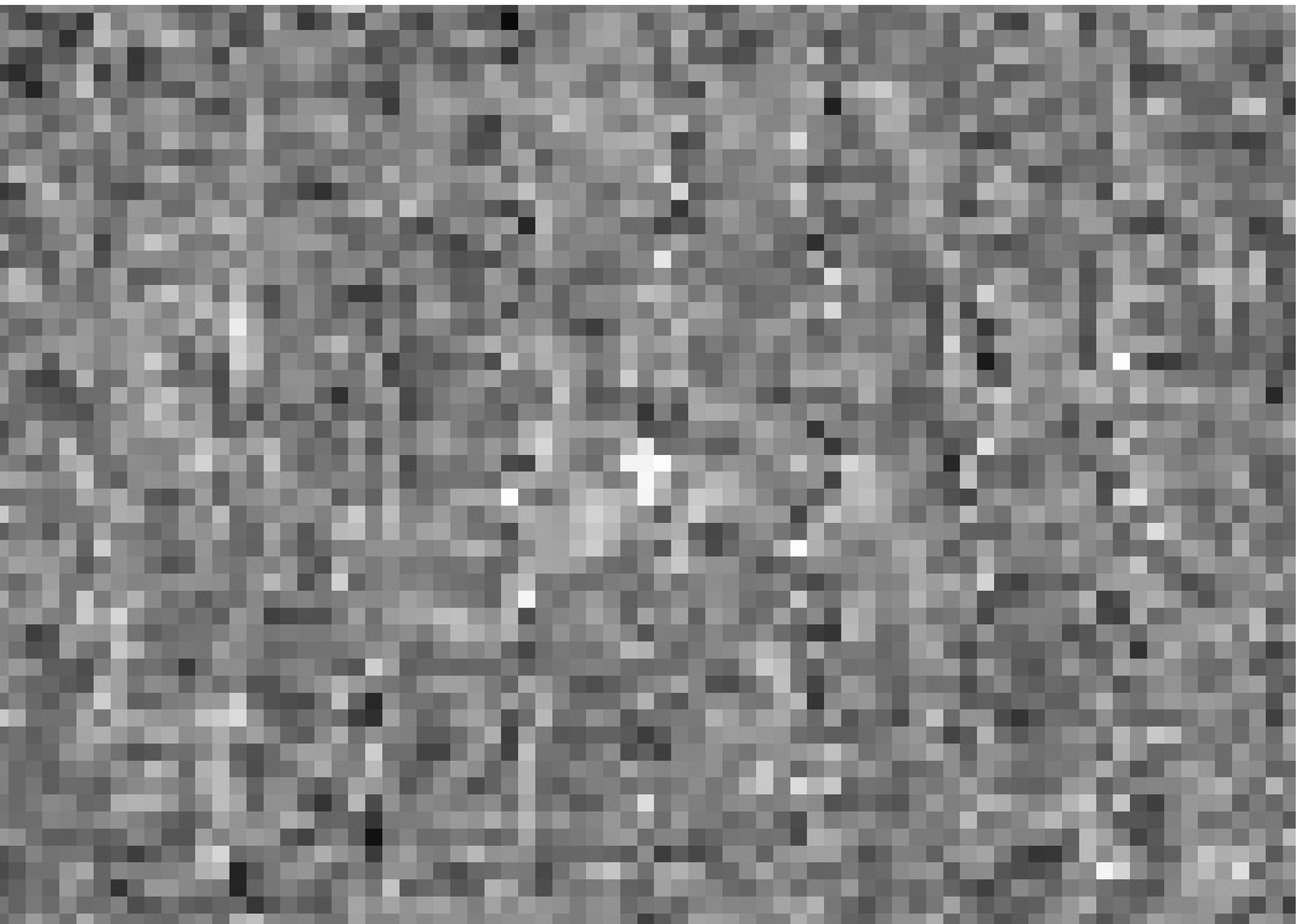} \hfil \includegraphics[width=0.45\columnwidth]{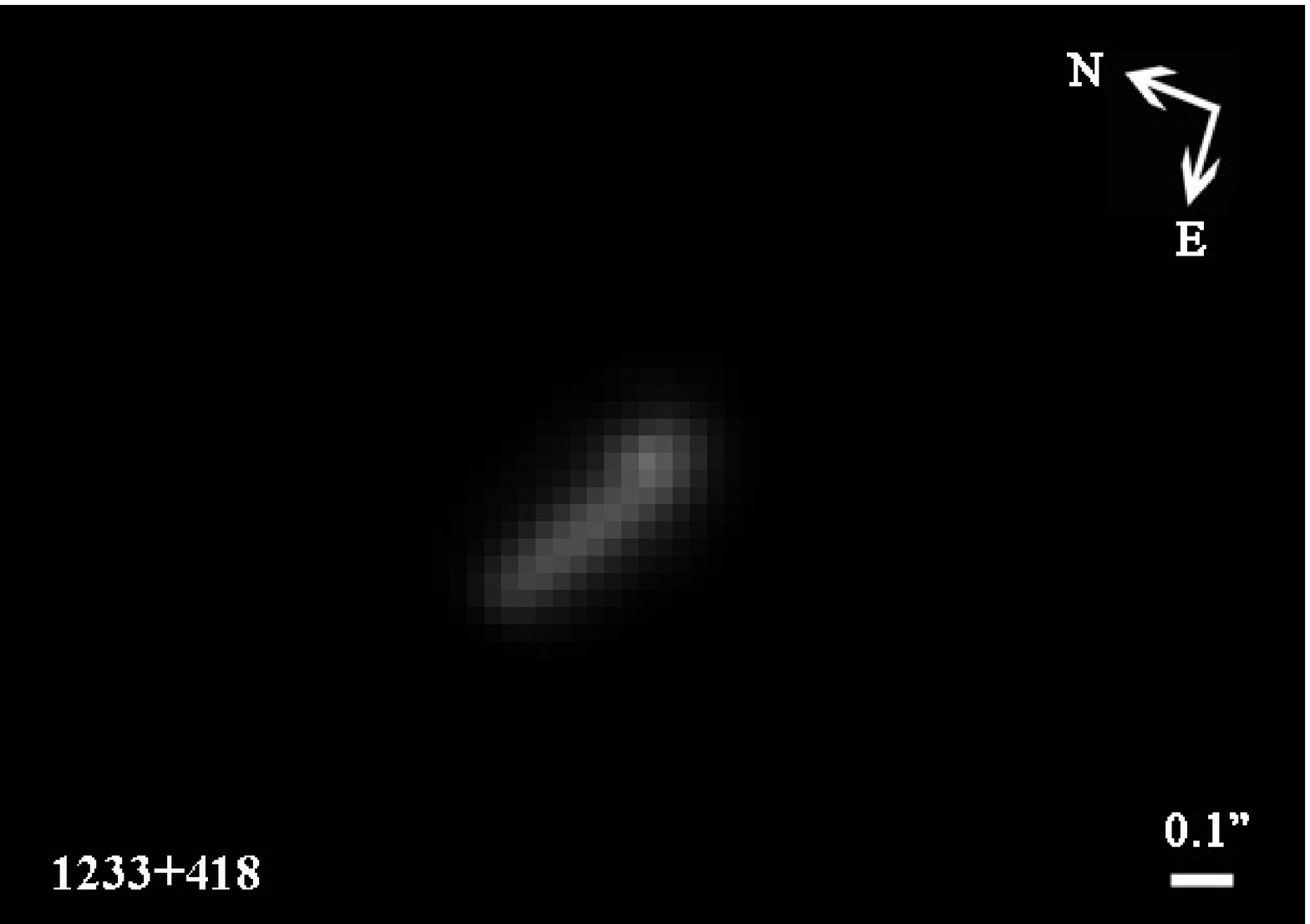} \hfil \includegraphics[width=0.45\columnwidth]{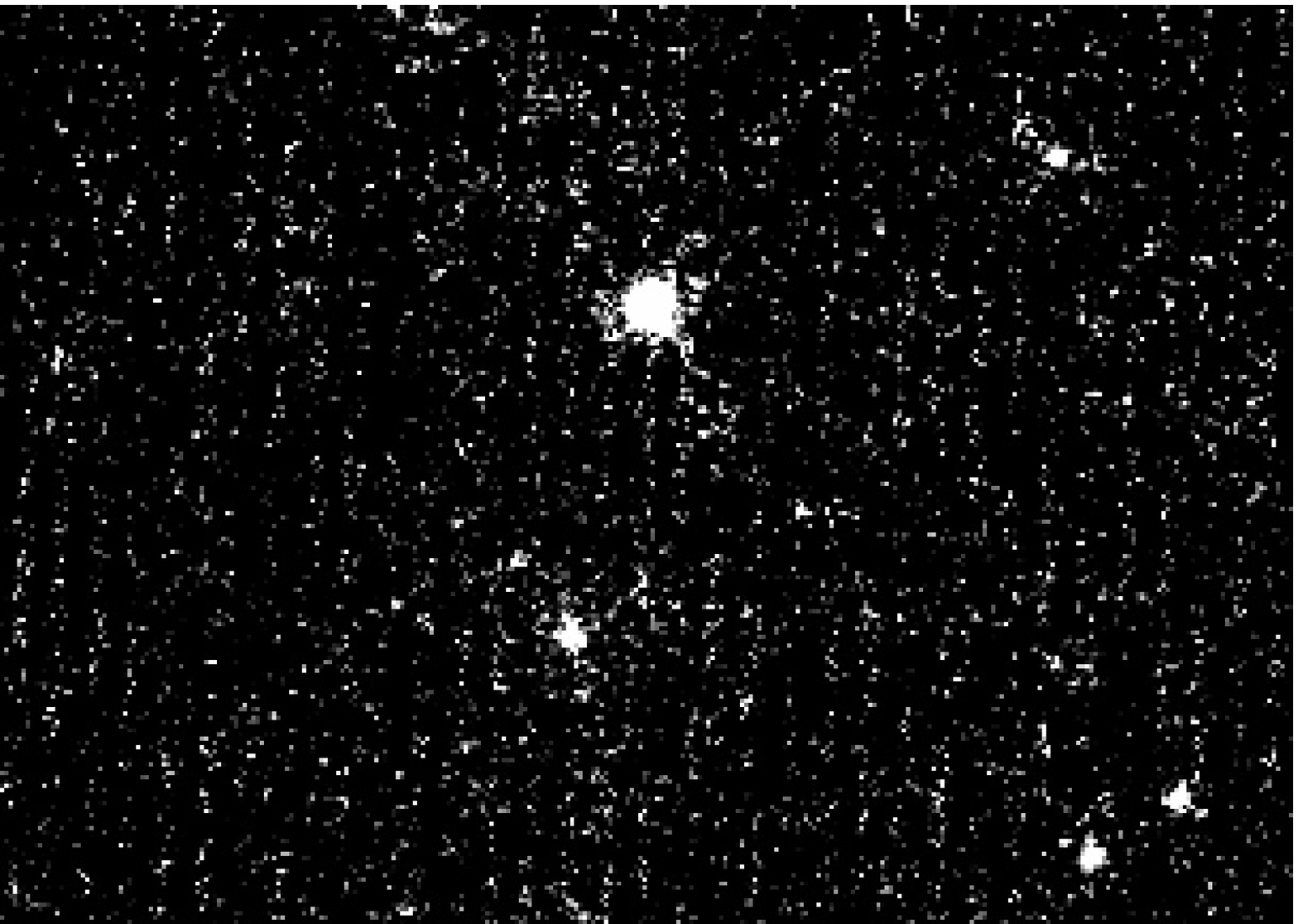} \hfil \includegraphics[width=0.45\columnwidth]{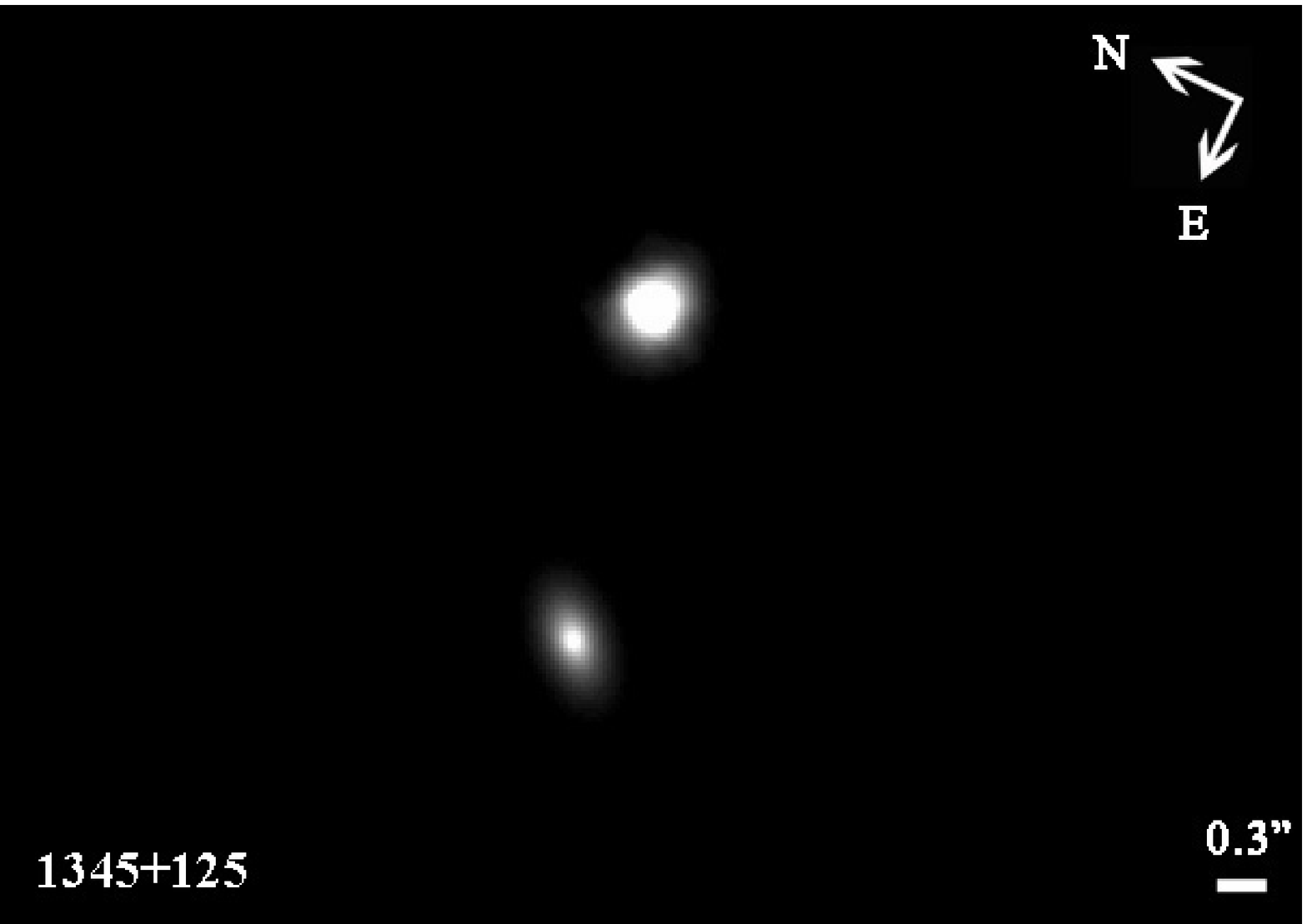} 
\includegraphics[width=0.45\columnwidth]{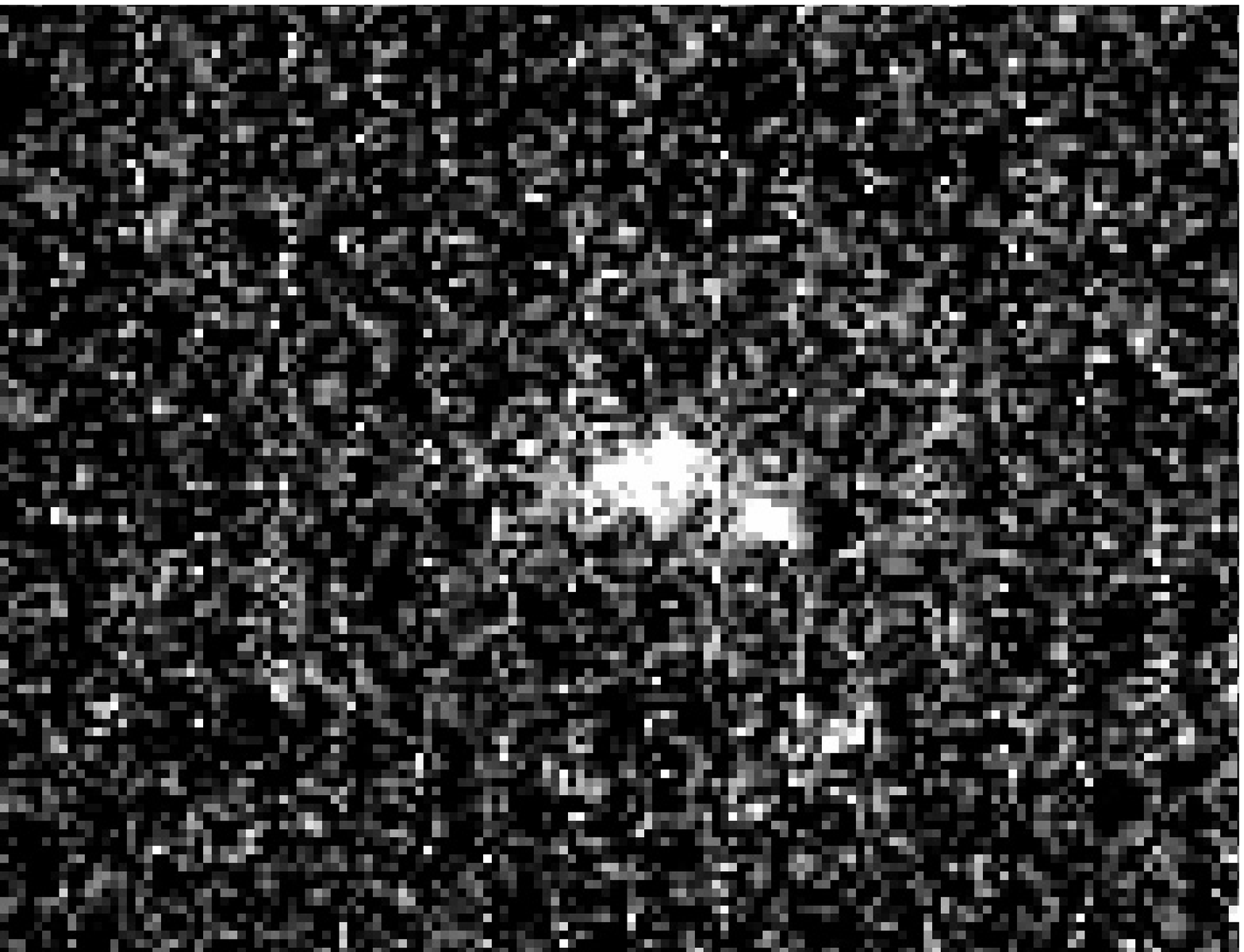} \hfil \includegraphics[width=0.45\columnwidth]{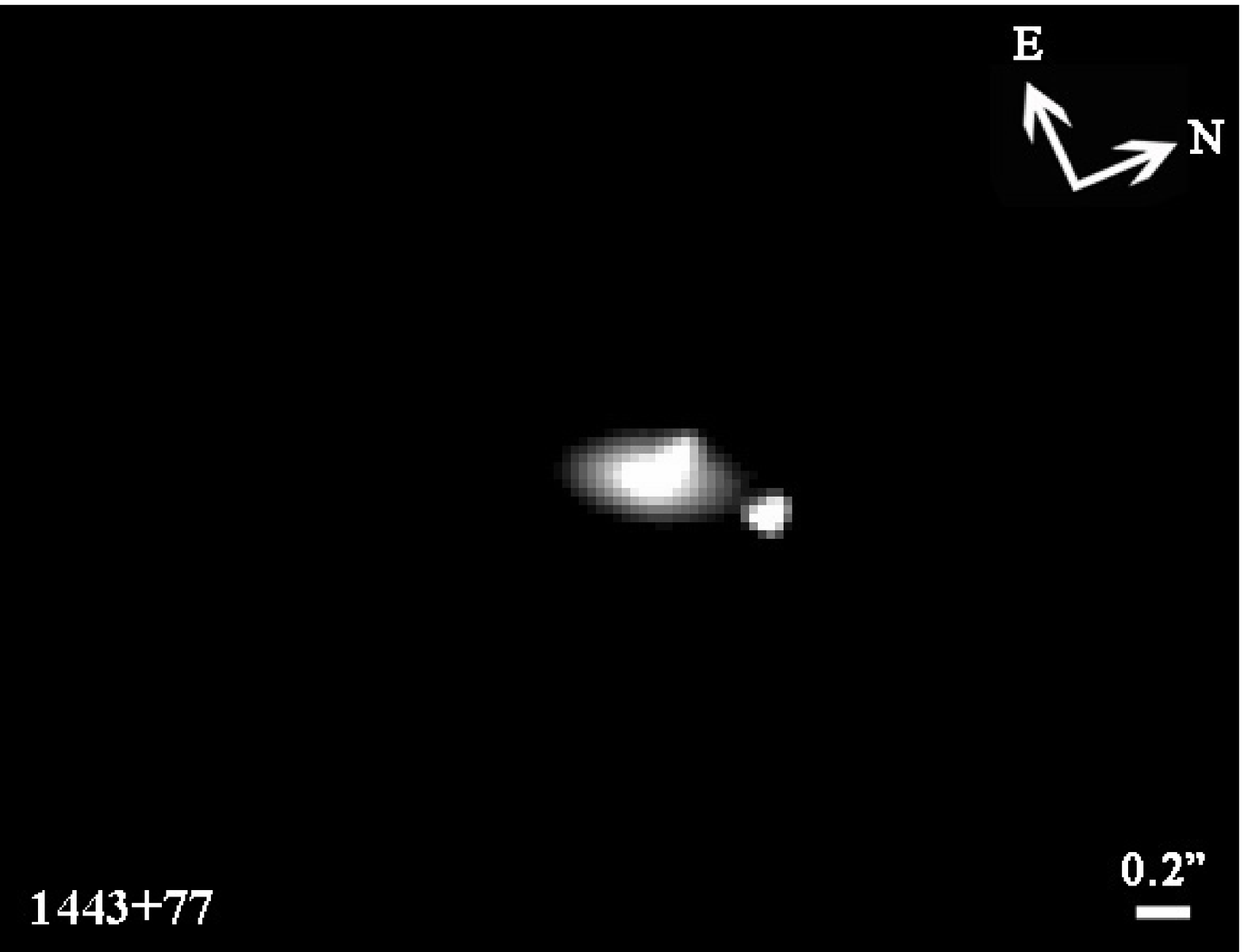} \hfil \includegraphics[width=0.45\columnwidth]{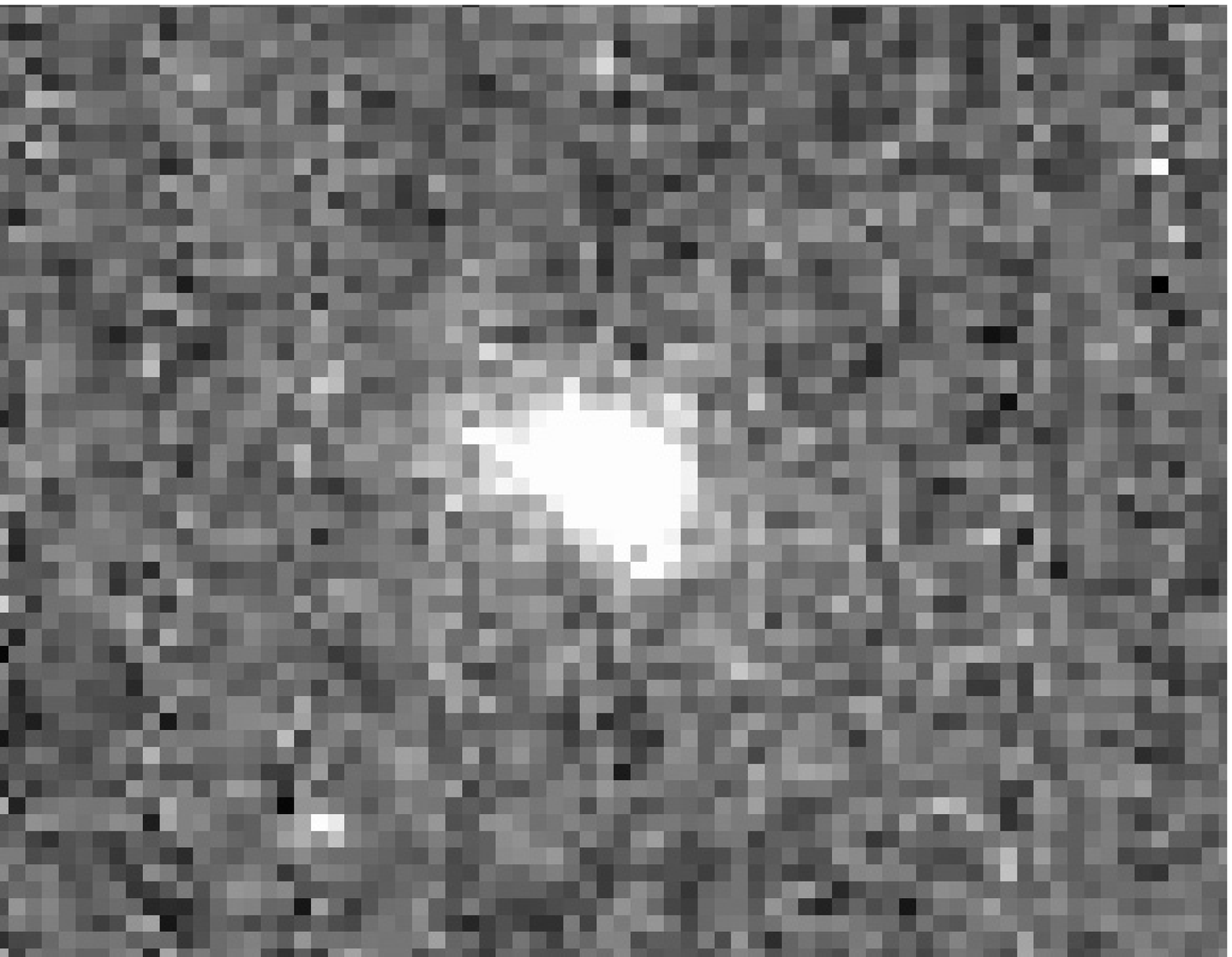} \hfil \includegraphics[width=0.45\columnwidth]{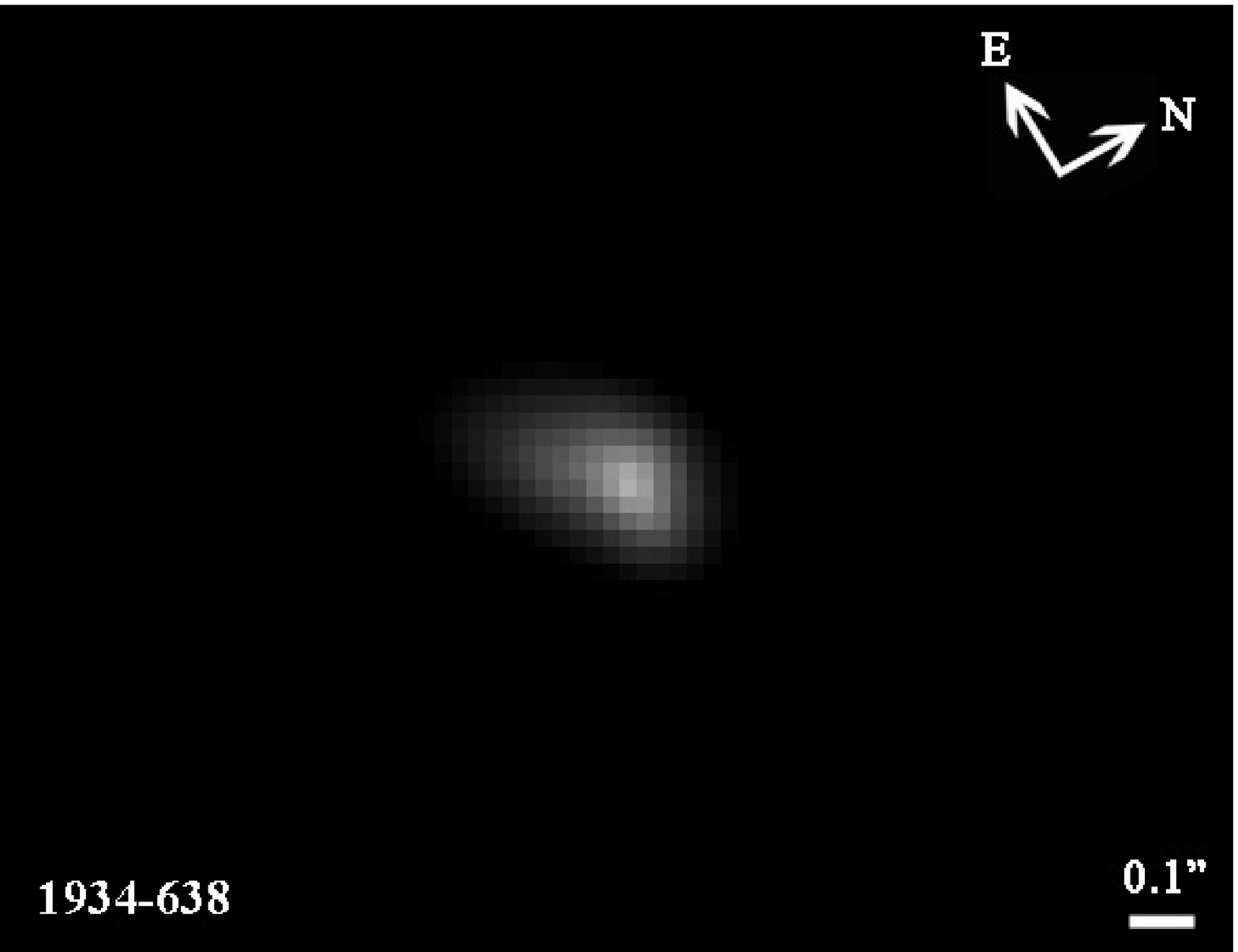} 
\includegraphics[width=0.45\columnwidth]{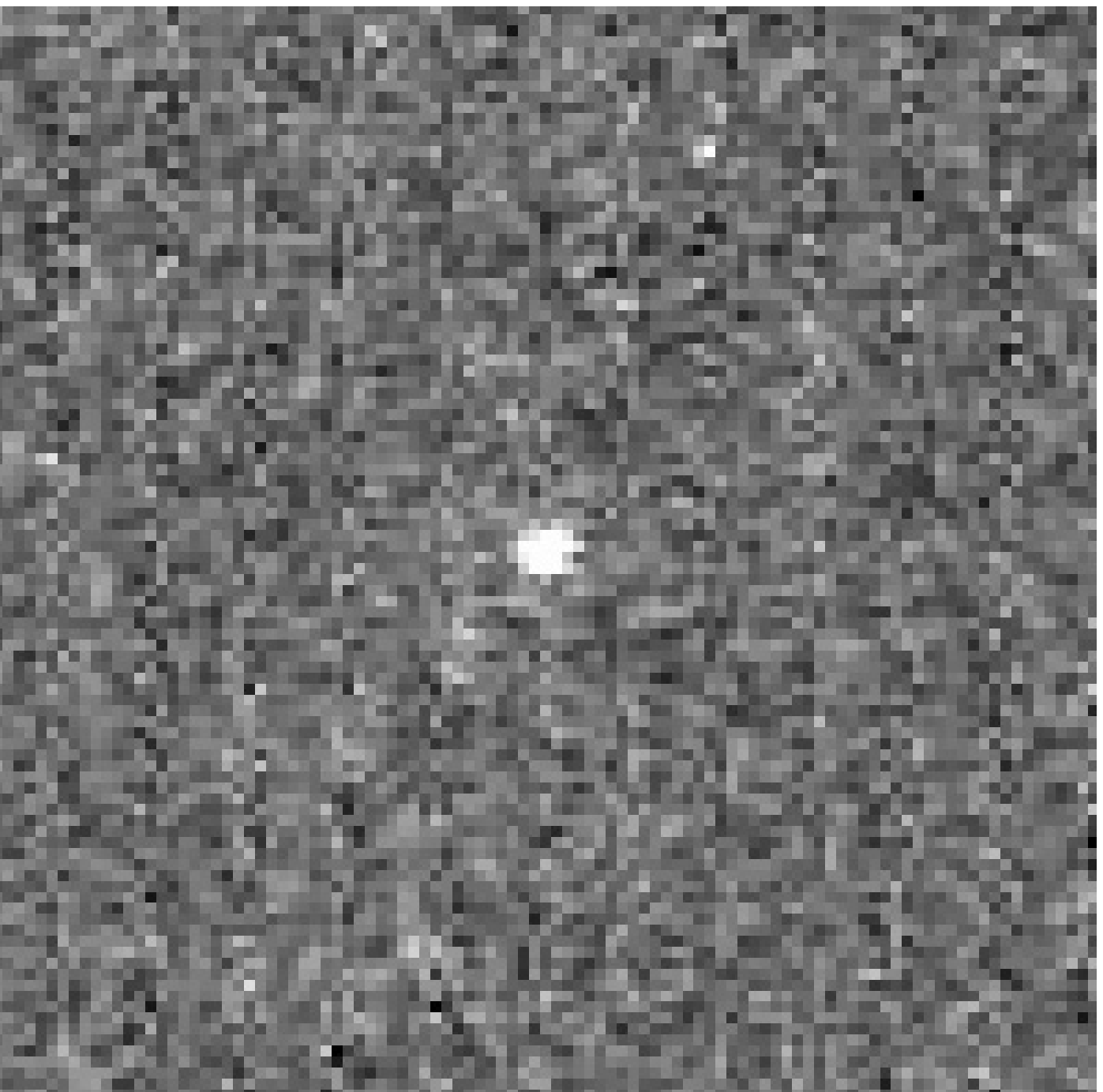} \hfil \includegraphics[width=0.45\columnwidth]{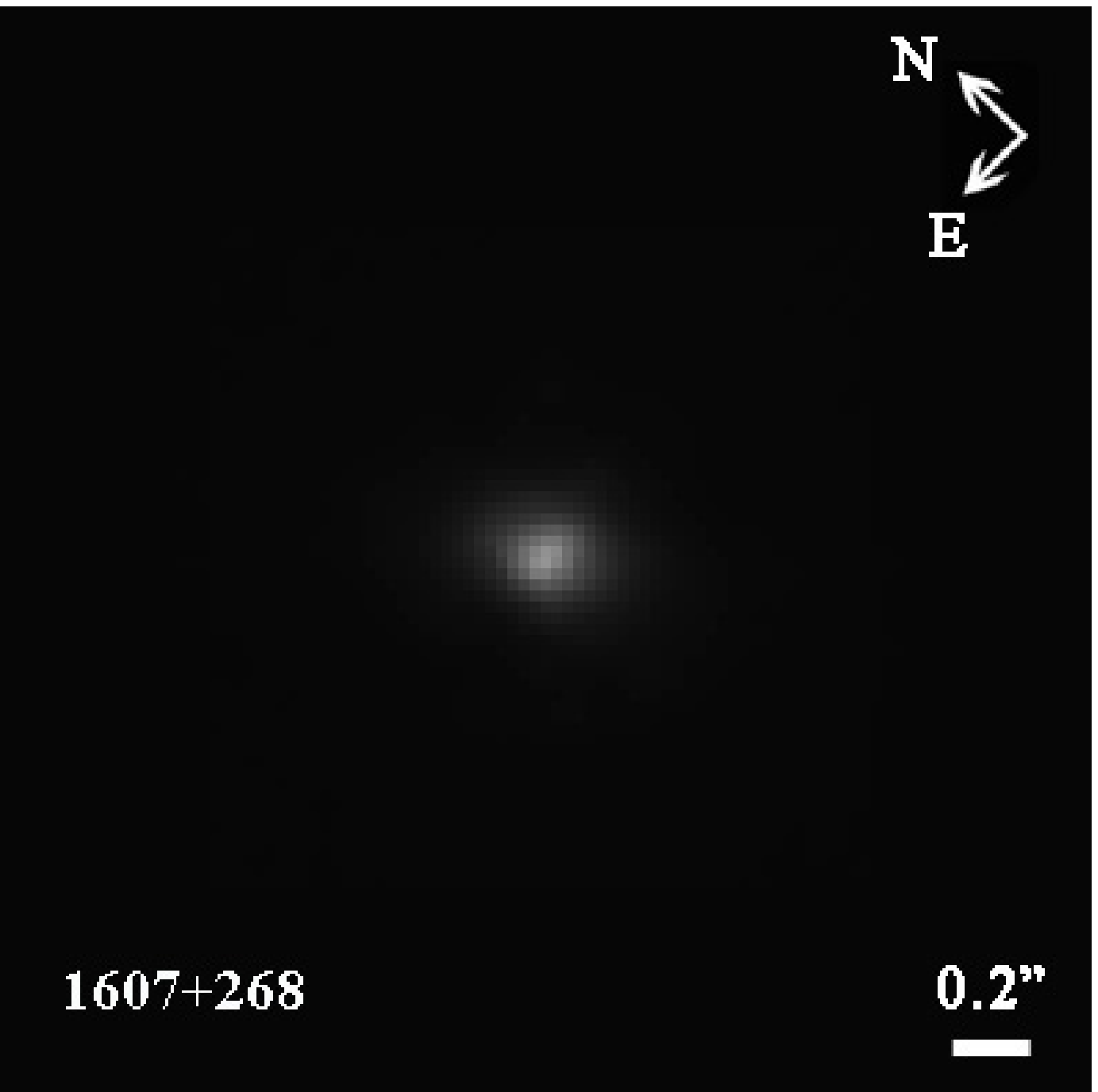} \hfil \includegraphics[width=0.45\columnwidth]{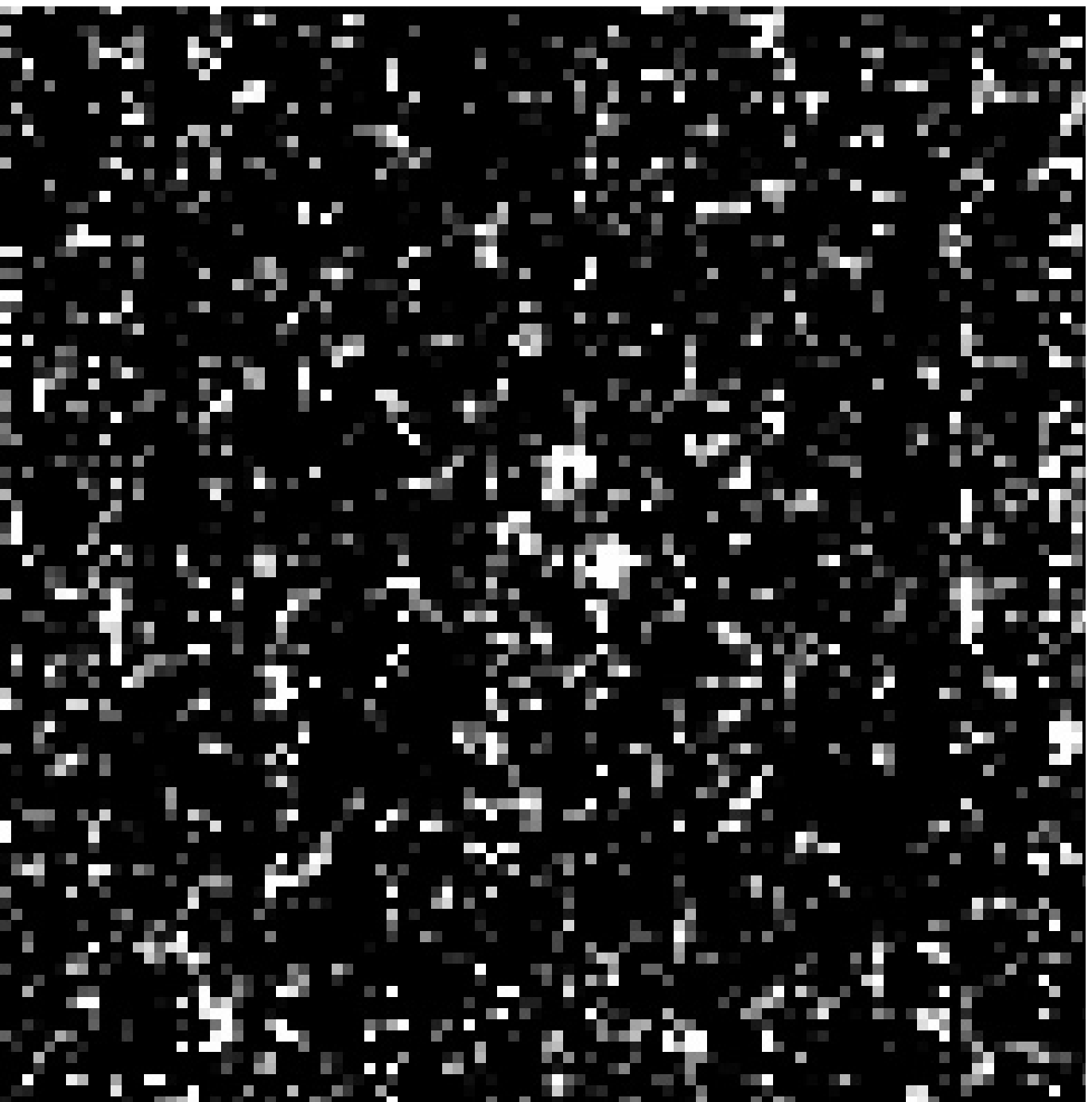} \hfil \includegraphics[width=0.45\columnwidth]{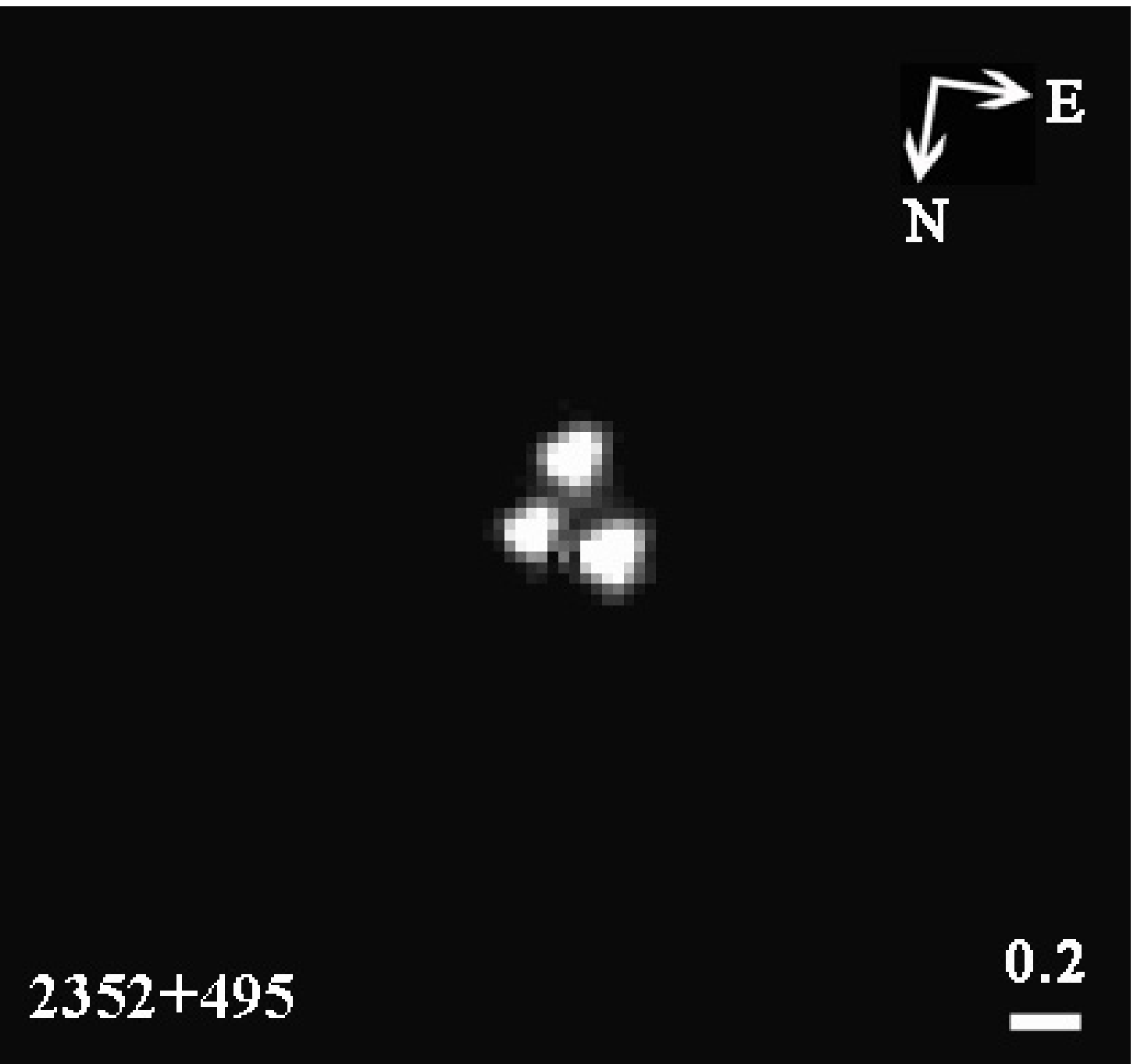} 

\caption{UV images (left panels) and GALFIT models (right panels). Only sources with complex morphologies in the UV image are shown. \label{galfigs}}
\end{center}
\end{figure*}